\documentclass[aps,pra,showpacs,twocolumn,groupedaddress]{revtex4}

\usepackage{graphicx}
\usepackage{amsmath}
\usepackage{amsfonts}
\usepackage{amssymb}
\usepackage{epsfig}
\usepackage{bm}

\begin{document}

\title{Ground State Geometry  of Binary Condensates in 
       Axisymmetric Traps}

\author{S. Gautam and D. Angom}
\affiliation{Physical Research Laboratory,
         Navarangpura, Ahmedabad - 380 009\\}

\date{\today}

\begin{abstract}
   We show that the ground state interface geometry of binary condensates in 
the phase separated regime undergoes a smooth transition from planar to 
ellipsoidal to cylindrical geometry. This occurs for condensates with 
repulsive interactions as the trapping potential is changed from prolate to 
oblate. The correct ground state geometry emerges when the interface energy 
is included in the energy minimization. Where as energy minimization based 
on Thomas-Fermi approximation gives incorrect geometry. 
\end{abstract}


\maketitle


\section{\label{section-1}Introduction}
Two species Bose-Einstein condensate (TBEC), consisting of two different 
hyperfine spin sates of $^{87}$Rb, was first observed by Myatt et al 
\cite{Myatt}. Since then, TBECs of different atomic species ($^{41}$K 
and $^{87}$Rb) \cite{Modugno} and of different isotopes of the same atomic 
species \cite{Papp} have been experimentally realized. This has led to several 
experimental and theoretical investigations on different aspects of TBECs.
The remarkable feature of TBECs which is absent in a single component 
Bose-Einstein condensates (BECs) is the phenomenon of phase separation. In 
Thomas-Fermi approximation (TFA), the phase separation occurs when
all the inter atomic interactions are repulsive and the inter species 
repulsion exceeds the geometric mean of the intra species repulsive 
interactions.
    
  Depending upon the properties of the condensates and trapping potential
parameters, the ground state of TBECs assumes a configuration which minimizes 
the total energy. The structure of the ground state plays an important role 
in dynamical phenomena like Rayleigh–Taylor \cite{Gautam,Sasaki}
instability, Kelvin-Helmholtz instability \cite{Takeuchi}, pattern 
formation at the interface \cite{Saito} etc. It was recently demonstrated that
quantum Rayleigh–Taylor instability can be observed in a very controlled 
way with TBECs in cigar shaped traps \cite{Gautam}.

 In the phase separated regime, the interface energy of the two component 
species defines the geometry of the ground state. In a previous work, the 
ground state geometry of TBECs was examined within the TFA \cite{Ho} that is 
without the interface energy. In later works, \cite{Timmermans,Ao,Barankov}
the contribution from the interface energy was incorporated. From these it is 
observed that the analytic approximations for interface energy are not 
sufficient enough to explain the experimental results of strongly 
segregated ground  states \cite{Papp}. A recent work \cite{Schaeybroeck}  
reported a more accurate determination of the interface energy. It
explains the stationary state geometry of the strongly segregated TBECs 
more precisely.   
    
 In this paper we provide a semi-analytic scheme to determine the stationary 
state structure of TBEC in axisymmteric traps. For this, we follow the 
ansatz adopted in Ref. \cite{Trippenbach} i.e. to minimize the total energy 
of TBEC with fixed number of particles of each species in TFA. In 
section \ref{section-2} of the manuscript, we identify three geometries
which a TBEC can assume depending on the trapping potential parameters. Based 
on TFA, we provide a semi-analytic scheme to determine the stationary state 
parameters of the ground state for each of these three geometries. In 
section \ref{section-3}, we analyze the crucial role played by the 
interface energy in determining the ground state structure of the TBEC.


\section{TBEC in Axisymmetric Traps}
\label{section-2}
We consider TBEC in axisymmetric trapping potentials
\begin{equation}
   V_i(r,z) = \frac{m_i\omega^2}{2}(\alpha_i^2r^2 + \lambda_i^2 z^2),
   \label{eq.pots}
\end{equation}
where $i=1,2$ is the species index, and $\alpha_i$ and $\lambda_i$ are the 
anisotropy parameters. In the mean field approximation, the stationary state 
solution of binary condensate is described by a set of coupled 
Gross-Pitaevskii equations
\begin{eqnarray}
   &&\left[ \frac{-\hbar^2}{2m_i}\nabla^2 + V_i(r,z) + U_{ii}|\psi_i(r,z)|^2 
	+ U_{ij}|\psi_{j}(r,z)|^2 \right] \nonumber \\
   &&\psi_{i}(r,z) =\mu_i\psi_{i}(r,z), 
  \label{eq.gp}
\end{eqnarray}
here $i$ and $j=3-i$ are species indices; $U_{ii} = 4\pi\hbar^2a_i/m_i$ with 
$m_i$ as mass and $a_i$ as s-wave scattering length, is the intra-species 
interaction; $U_{ij}=2\pi\hbar^2a_{ij}/m_{ij}$ with 
$m_{ij}=m_i m_j/(m_i+m_j)$ as reduced mass and $a_{ij}$ as inter-species 
scattering length, is the inter-species interaction term and $\mu_i$ is 
the chemical potential of the $i$th species.

  When the number of atoms are large, the TFA is applicable to obtain
the stationary state solutions of Eq.(\ref{eq.gp}). In this limit
the kinetic energy is neglected in comparison to interaction energy. We 
consider the interaction parameter $U_{12}>\sqrt{U_{11}U_{22}}$, such that 
the two components are phase separated. That is, the two components occupy 
different regions of the trapping potentials. Neglecting the overlap between 
the species, stationary state solutions within TFA are
\begin{equation}
\label{eq.tf_sol}
 |\psi_i(r,z)|^2 = \frac{\mu_i-V_i(r,z)}{U_{ii}}.
\end{equation}
where $\mu_i$ is fixed by the number of atoms of the corresponding species.

\begin{figure}[h]
   \includegraphics[width=8cm,angle=0]{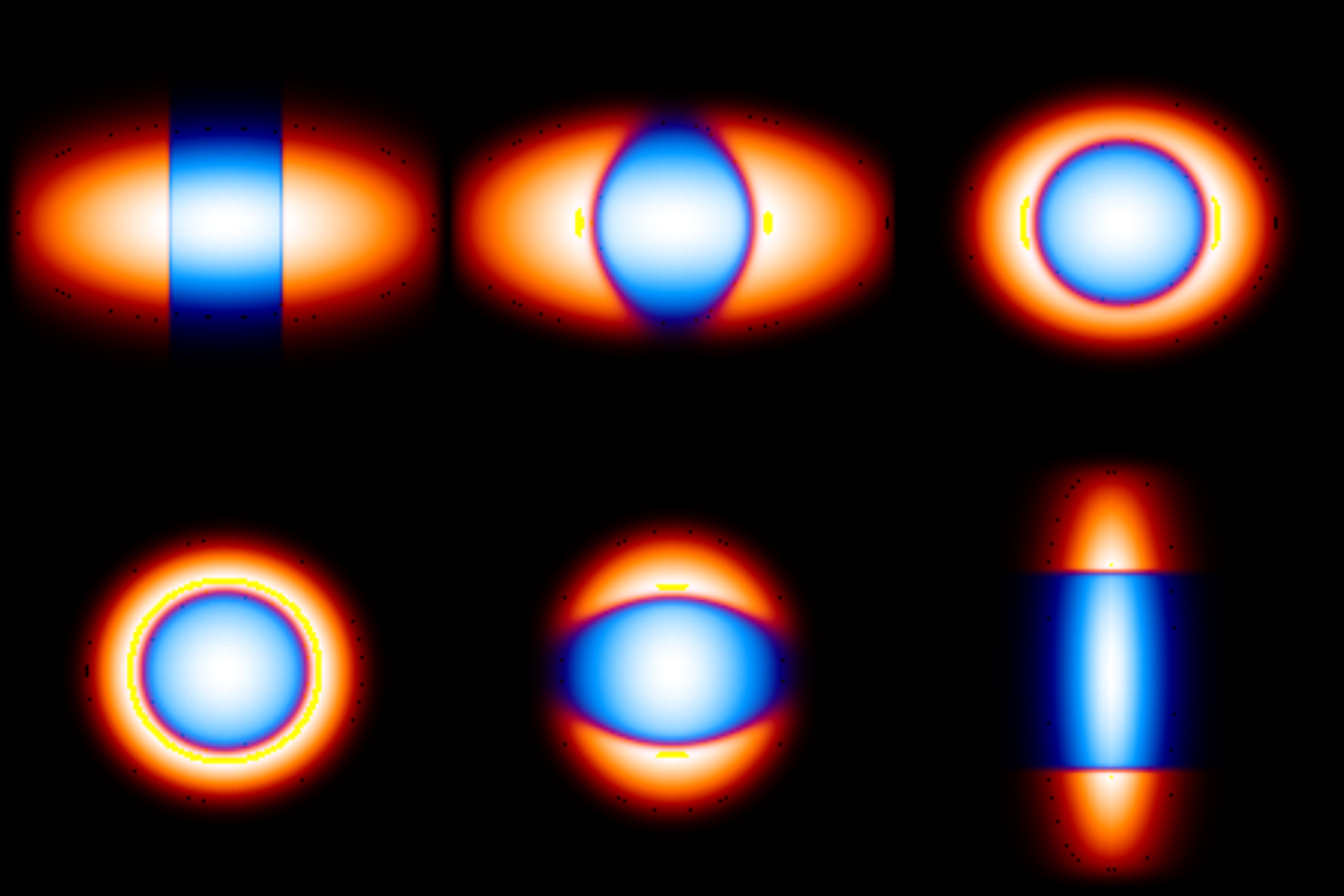}
   \caption{\label{fig.tbec_phase_images} The false color coded images of
             $|\psi_i(r, z)|$, for TBEC consisting of $^{85}$Rb (inner 
             component) and $^{87}$Rb (outer component), on $r-z$ plane with 
             vertical and the horizontal directions corresponding to radial 
             and axial coordinate respectively. The wave functions are 
             obtained by numerically solving Eq.(\ref{eq.gp}) using 
             $a_{\rm osc}$, $a_i$ and $a_{12}$ taken from Ref. \cite{Papp},
             referred to as {\em set a} in the text. The images correspond to 
             different values of $\lambda_i$ but same $N_i=50,000$. In the 
             first row, starting from left ($\lambda_1$, $\lambda_2$) are 
             (0.020, 0.022), (0.50, 0.50) and (0.85,0.85). While the second 
             row corresponds to (1.0, 1.0), (1.5, 1.5) and (50.0, 50.0). 
             As is evident the interface geometry changes continuously from 
             planer to ellipsoidal to cylindrical.}
\end{figure}

 The total energy of the TBEC in the phase separated regime is
\begin{eqnarray}
   E  & = & \int dV\left[ V_1(r,z)|\psi_1(r,z)|^2 + V_2(r,z)|\psi_2(r,z)|^2 
	        + \right.    \nonumber    \\
      &   & \left.\frac{1}{2}U_{11}|\psi_1(r,z)|^4 + \frac{1}{2}U_{22}|
            \psi_2(r,z)|^4 \right] .
   \label{eq.en}
\end{eqnarray}
Depending upon the anisotropy parameters, the TBEC can have three distinct
spatial distributions in axisymmetric traps. The distinguishing feature of 
these structures is the geometry of the interface, which can be planar, 
cylindrical or ellipsoidal. The smooth transition of interface geometry, for 
the TBEC of $^{85}$Rb-$^{87}$Rb mixture, from planar to ellipsoidal and 
finally to cylindrical is shown in Fig. \ref{fig.tbec_phase_images}. These 
features are most prominent when the TBEC is  strongly segregated and for the 
detailed examination of our scheme we choose $^{85}$Rb-$^{87}$Rb experiments 
of Papp et al \cite{Papp}. Where two of the geometries, planar and ellipsoidal, 
were observed.


\subsection{Planar interface}
It has been observed experimentally \cite{Papp} that in cigar shaped traps
($\lambda_i\ll\alpha_i$) the TBEC assumes a sandwich structure with 
planar interface between the two species. In this structure the phase 
separation occurs along the axial direction, with the weakly interacting 
component sandwiched by the strongly interacting one. There are two 
realizations of this: coincident and shifted trapping potentials.

\subsubsection{Coincident trap centers}
 An idealized choice of $V_i$ is with coincident centers. If $z=\pm L_1$ are 
the locations of the planes separating the two components and $L_2$, the axial 
size of binary condensate. Then the problem of determining the structure of 
the  TBEC is equivalent to calculating $L_1$. If $N_i$ and $R_i$ are the 
number of atoms and radial size of the $i$th species respectively, then 
\begin{eqnarray}
  N_i = 2\pi\int_0^{R_i}r dr \int_{-L_i}^{L_i}dz |\psi_i(r,z)|^2.
  \label{eq.ni}
\end{eqnarray}
From Eq.(\ref{eq.tf_sol}), we get
\begin{eqnarray}
  \label{eq.mu}
  N_1 & = & 2 \pi  \left(\frac{\omega ^2 L_1^5 m_1 \lambda _1^4}{20 U_{11} 
          \alpha _1^2}-\frac{L_1^3 \lambda _1^2 \mu _1}{3 U_{11} \alpha _1^2}+
          \frac{L_1 \mu _1^2}{\omega ^2 m_1 U_{11} \alpha _1^2}\right),
                        \nonumber\\
  N_2 & = & 4\pi\left(\frac{4 \sqrt{2} \mu _2^2 \sqrt{\frac{\mu _2}
          {\omega ^2 \lambda _2{}^2 m_2}}}{15 \omega ^2 m_2 U_{22} 
          \alpha _2^2}+\frac{ L_1}{120 U_{22} \alpha _2^2}\left(5 
          \omega ^2 L_1^4 m_2 \lambda _2^4\right.\right.\nonumber\\
      &   &\left.\left.-\frac{8 \omega ^2 m_2 \left(L_1^2 \lambda _2^2\right)
           {}^{5/2}}{\lambda _2 L_1} +20 L_1^2 \lambda _2^2 \mu _2-
           \frac{60 \mu _2^2}{\omega ^2 m_2}\right)\right).\nonumber\\
\end{eqnarray}

Similarly, the total energy in Eq.(\ref{eq.en}) is
\begin{eqnarray}
   E & = & \frac{4\pi}{1680 \omega ^2 m_2 U_{22} \alpha _2^2}
           \bigg(-21 \omega ^6 L_1^7 m_2^3 \lambda _2^6 + 
           16 \omega ^6 L_1^7 m_2^3 \lambda _2^6 +\nonumber\\
     &   &112 \omega ^4 L_1^5 m_2^2 \lambda _2^4 \mu _2 -
          112 \omega ^4 L_1^5 m_2^2 \lambda _2^4 \mu _2+	 
          140 \omega ^2 L_1^3 m_2 \lambda _2^2 \mu _2^2   \nonumber\\
     &   &-560 L_1 \mu _2^3+320 \sqrt{2} \mu _2^3 
          \sqrt{\frac{\mu _2}{\omega ^2 \lambda _2{}^2 m_2}}\bigg)+
					\nonumber\\
     &   & 2 \pi  \left(\frac{\omega ^4 L_1^7 m_1^2 
           \lambda _1^6}{168 U_{11} \alpha _1^2}- \frac{L_1^3 \lambda _1^2 
           \mu _1^2}{6 U_{11} \alpha _1^2} +
           \frac{2 L_1 \mu _1^3}{3 \omega ^2 m_1 U_{11} \alpha _1^2}\right).
\label{eq.E}
\end{eqnarray}
Here, $L_1$ is determined by variational minimization of $E$ 
with $L_1$  as the variational parameter and constraints that $\mu_1$, and 
$\mu_2$ satisfy Eq.(\ref{eq.mu}) for fixed $N_i$. In the constraint 
equations, we invert the expression of $N_1$ and obtain $\mu_1 $ as a 
function of $L_1$.  However, inverting $N_2$ to calculate $\mu_2$ is 
nontrivial. Hence, we implement the minimization numerically.

As mentioned earlier we consider the TBEC of $^{85}$Rb-$^{87}$Rb with
$N_i=50,000$. The scattering lengths $a_1=51a_o$, $a_2=99a_o$ and 
$a_{12}=214a_o$ are from the experimental results of Wieman and collaborators 
\cite{Papp}. Like wise the anisotropy parameters and trap frequency are 
$\alpha_i=1$, $\lambda_1=2.9/130$, $\lambda_2=2.6/130$ and $\omega=130$Hz 
respectively. From here on this choice of parameters is referred as the 
{\em set a}. For these parameters, the minima 
of $E$ occurs at $32.3a_{\rm osc}$. This is in very good agreement with 
the value of $33.8a_{\rm osc}$ obtained from the numerical solution 
\cite{Muruganandam} of Eq.(\ref{eq.gp}). The contour plots showing the
absolute value of wave functions of $^{85}$Rb and $^{87}$Rb, obtained by
numerically solving Eq.(\ref{eq.gp}), are shown in first image from
left in upper panel of Fig. \ref{fig.tbec_phase_images}.

\begin{figure}[t]
   \includegraphics[width=8cm,angle=0]{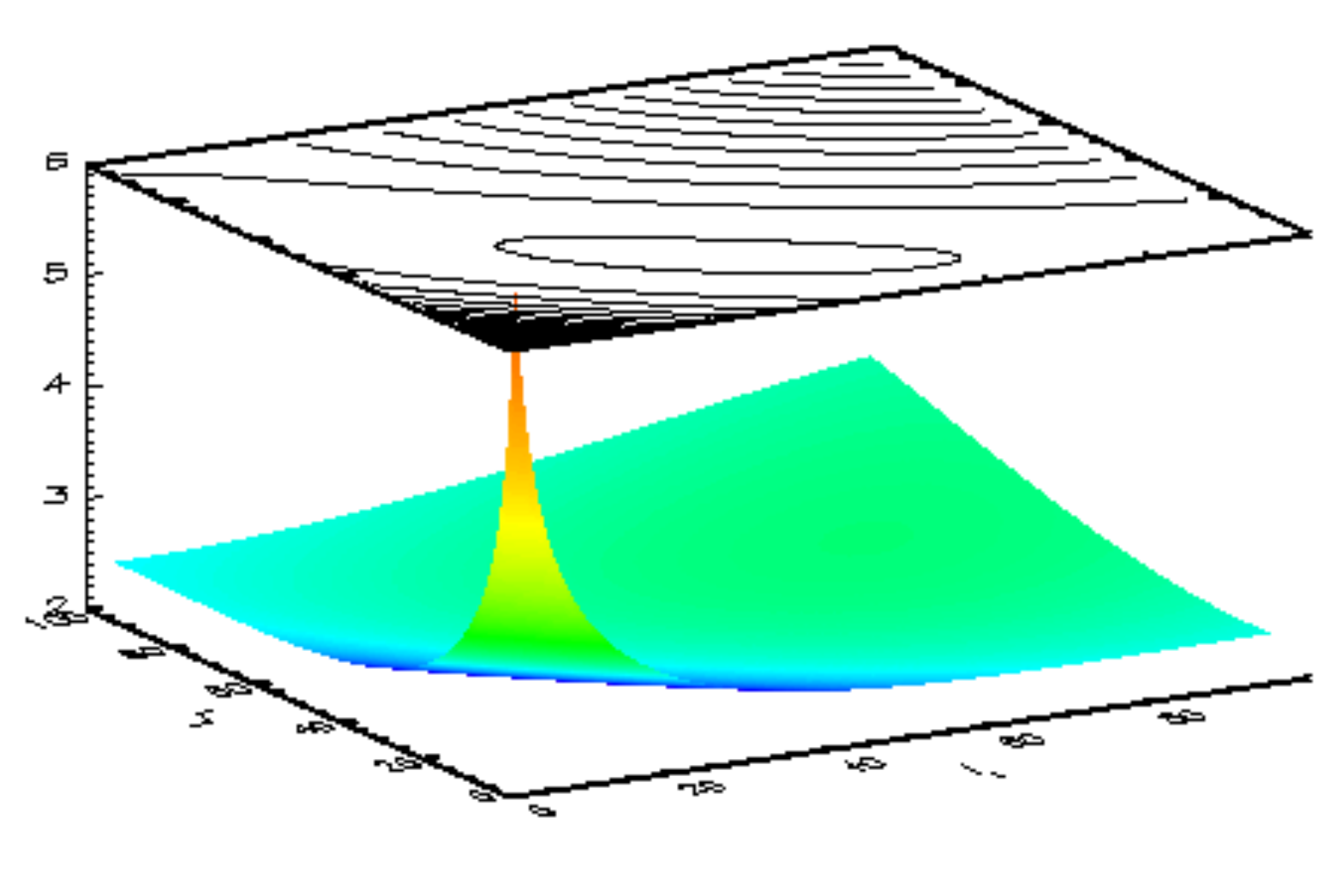}
   \caption{\label{fig.cig2} Surface and contour plot of the variation of 
            $E$ as function of $l_1$  and $L_1$. The minima of $E$ occurs
            at $l_1=39.5a_{\rm osc}$ and $L_1=26.0a_{\rm osc}$. These are 
            close to $l_1=40.15a_{\rm osc}$ and $L_1=27.95a_{\rm osc}$ 
            obtained from the numerical solution of coupled GP equations.}
\end{figure}

\subsubsection{Separated trap centers}
In the experimental realizations, the gravitational potential and tilts
in the external field configurations tends to separate the minima of the 
effective potentials. Normally, in cigar shaped traps, the tilt angle is 
small and separation is effectively along the axial direction. Then 
potentials with separation $z_0$ are
\begin{eqnarray}
   V_1(r,z) &= & \frac{m_1\omega^2}{2}(\alpha_1^2r^2 + \lambda_1^2 z^2),
                             \nonumber\\
   V_2(r,z) &= & \frac{m_2\omega^2}{2}\left [ \alpha_2^2r^2 + 
                 \lambda_2^2 (z-z_0)^2\right ].
   \label{eq.pots_shift}
\end{eqnarray}
Due to the lack of axial symmetry, $z=-l_1$ and $z=L_1$ are the two planes 
separating the two components. These, $l_1$ and $L_1$, are the 
parameters to minimize $E$. Like in the previous subsection $N_1$, $N_2$
and $E$ can be evaluated and are presented in the appendix.
For the parameter {\em set a} and $z_0=3.4\mu$m, the minima of $E$ occurs 
when  $l_1$ and $L_1$ are $39.5a_{\rm osc}$ and $26.0a_{\rm osc}$ respectively.
The overall trend of $E$ is shown in Fig. \ref{fig.cig2}.


\subsection{Ellipsoidal Interface}
 As the anisotropy parameter $\lambda$ is increased, beyond a critical value 
$\lambda_a$ the interface geometry change from planer to ellipsoidal. Where 
one species envelopes the other. This is the preferred interface geometry,
for the phase separated TBEC in axisymmteric traps, without interface energy.
Consider trapping potentials with coincident centers, if $R_i$ and $L_i$ are 
equatorial (along radial direction) and polar (along axial direction) radii
of the $i$th species respectively, then
\begin{equation}
   N_i = 2\pi\int_0^{R_i}r dr \int_{-L_i}^{L_i}dz |\psi_i(r,z)|^2.
 \label{eq.Ni}
 \end{equation}
From Eq.(\ref{eq.tf_sol}) and Eq.(\ref{eq.en}), we get
\begin{eqnarray}
N_1 & = &\frac{-2\pi R_1\alpha_1(3\omega^2m_1R_1^4\alpha_1^2-10R_1^2\mu_1)}
        {15U_{11}\lambda_1},\\
N_2 & = & \frac{2\pi}{15U_{22}\lambda_1^3}\Bigg(\omega^2m_2R_1^5\alpha_1
        (2\alpha_2^2\lambda_1^2 +\alpha_1^2\lambda_2^2)\nonumber\\
    &   & -10R_1^3\alpha_1\lambda_1^2\mu_2+
		    \frac{8\sqrt{2}\lambda_1^3\mu_2^{5/2}}
		    {\lambda_2\alpha_2^2\omega^3m_2^{3/2}}\Bigg),\\
E & = & \frac{\pi}{210\omega^2m_2U_{11}U_{22}\alpha_2^2\lambda_1^5}
        \left(-15\omega^6m_1^2m_2R_1^7U_{22}\alpha_1^5\alpha_2^2\lambda_1^4
				\right.\nonumber\\
  &   & +\omega^6m_2^3R_1^7U_{11}\alpha_1\alpha_2^2(8\alpha_2^4\lambda_1^4+
        4\alpha_1^2\alpha_2^2\lambda_1^2\lambda_2^2+3\alpha_1^4\lambda_2^4)
				 \nonumber\\
  &   &+160\sqrt{2}U_{11}\lambda_1^5\mu_2^3\sqrt{\frac{\mu_2}
	     {\omega^2m_2\lambda_2^2}}\nonumber\\
  &   &\left.+140\omega^2m_2R_1^3\alpha_1\alpha_2^2\lambda_1^4(U_{22}\mu_1^2- 
	     U_{11}\mu_2^2)\right).
 \label{gs_ellip}
\end{eqnarray}
In TFA, the profile of density $|\psi_i(r, z)|^2$ has the same ellipticity 
$e$ as that of the trapping potential, which is a function $\lambda$. Then,
$L_i$ is $\alpha_i R_i/\lambda_i$, further more in TFA $\mu_2$ constrains the 
value of $R_2$.  These reduce the variation parameter to only $R_1$. The 
energy $E$ is then minimized numerically to find the equilibrium geometry of 
the phase separated TBEC. To examine the scheme consider the 
$^{85}$Rb-$^{87}$Rb mixture  with parameter {\em set a} and coincident
trapping potentials, however take $\lambda_i$ as 1.5. Then the equilibrium 
geometry is ellipsoidal with an equitorial radius ($R_1$ ) of 
$3.72a_{\rm osc}$. This is in very good agreement with value 
$3.75a_{\rm osc}$ obtained from the numerical solution of GP equations.
The contour plots showing the absolute value of wave functions of 
$^{85}$Rb and $^{87}$Rb, obtained by numerically solving Eq.(\ref{eq.gp}), 
are shown in second image from left in lower panel of 
Fig. \ref{fig.tbec_phase_images}.


\subsection{Cylindrical interface}
  On further increase of $\lambda$, beyond another critical value 
$\lambda_b$, the equilibrium interface geometry is like a cylinder. Where 
the axis of the interface coincides with the polar axis of the trapping 
potentials. This occurs when $\lambda_i > \alpha_i$ i.e. in the oblate
condensates. Here, the phase separation is along radial direction and 
analogous to planar interface in cigar shaped condensates. If $\rho$ is the 
radius of the interface cylinder, then in TFA 
\begin{eqnarray}
N_1 & = & -\frac{4\pi(\omega^2\rho^2m_1\alpha_1^2-2\mu_1)^2\sqrt
        {-\rho^2\alpha_1^2+2\mu_1/(\omega^2m_1)}}{15\omega^2m_1U_{11}
				\lambda_1\alpha_1^2}\nonumber\\   
    &   & + \frac{16\pi\sqrt{2}\mu_1^{5/2}}{15U_{11}\lambda_1\alpha_1^2
		    (m_1\omega^2)^{3/2}},\nonumber\\
N_2	& = &\frac{4\pi(\omega^2\rho^2m_2\alpha_2^2-2\mu_2)^2\sqrt{-\rho^2
         \alpha_2^2+2\mu_2/(\omega^2m_2)}}{15\omega^2m_2U_{22}\lambda_2
				 \alpha_2^2},\nonumber\\
E & = & \frac{-4\pi}{15U_{11}}\left(\frac{-20\sqrt{2}\mu_1^{7/2}}
        {7\lambda_1\alpha_1^2(m_1\omega^2)^{3/2}}
      +\frac{(\omega^2\rho^2m_1\alpha_1^2-2\mu_1)^2}
			 {7\omega^2m_1\lambda_1\alpha_1^2}\right.\nonumber\\
	&   & \left.(\omega^2\rho^2m_1\alpha_1^2+5\mu_1)\sqrt{-\rho^2\alpha_1^2+
	      \frac{2\mu_1}{\omega^2m_1}}\right)+\nonumber\\
	&   &  \frac{\sqrt{-\rho^2\alpha_2^2+2\mu_2/(\omega^2m_2)}}{7U_{22}
	       \lambda_2}\left(\frac{4\pi\omega^4\rho^6m_2^2\alpha_2^4}{15}+
				 \right.\nonumber\\
	&   & \left.\frac{4\pi\omega^2\rho^4m_2\alpha_2^2\mu_2}{15}-
	      \frac{64\pi \rho^2\mu_2^2}{15}+\frac{16\pi\mu_2^3}
				{3\omega^2m_2\alpha_2^2}\right).
\end{eqnarray}
Above set of equations define the stationary state of TBEC in the oblate
shaped condensates. Like in the planar geometry, $\rho $ is the parameter of 
variation. To verify the scheme, we consider pan cake shaped 
($\lambda_i \gg \alpha_i$ ) TBEC of $^{85}$Rb-$^{87} $Rb mixture in
coincident traps with $\lambda_i $ as 50.0 and parameter {\em set a}. Then 
from our scheme the equilibrium state has cylindrical interface of radius
$5.84a_{\rm osc}$. The value from numerical solution of GP equation is 
$5.80a_{\rm osc}$. The two results are in very good agreement and validate 
our minimization scheme. The contour plots showing the absolute value of 
wave functions of $^{85}$Rb and $^{87}$Rb, obtained by numerically solving 
Eq.(\ref{eq.gp}), are shown in third image from left in lower panel of 
Fig.\ref{fig.tbec_phase_images}.


\section{Role of the interface energy}
\label{section-3}
\begin{figure}[h]
\includegraphics[height=8cm,angle=-90]{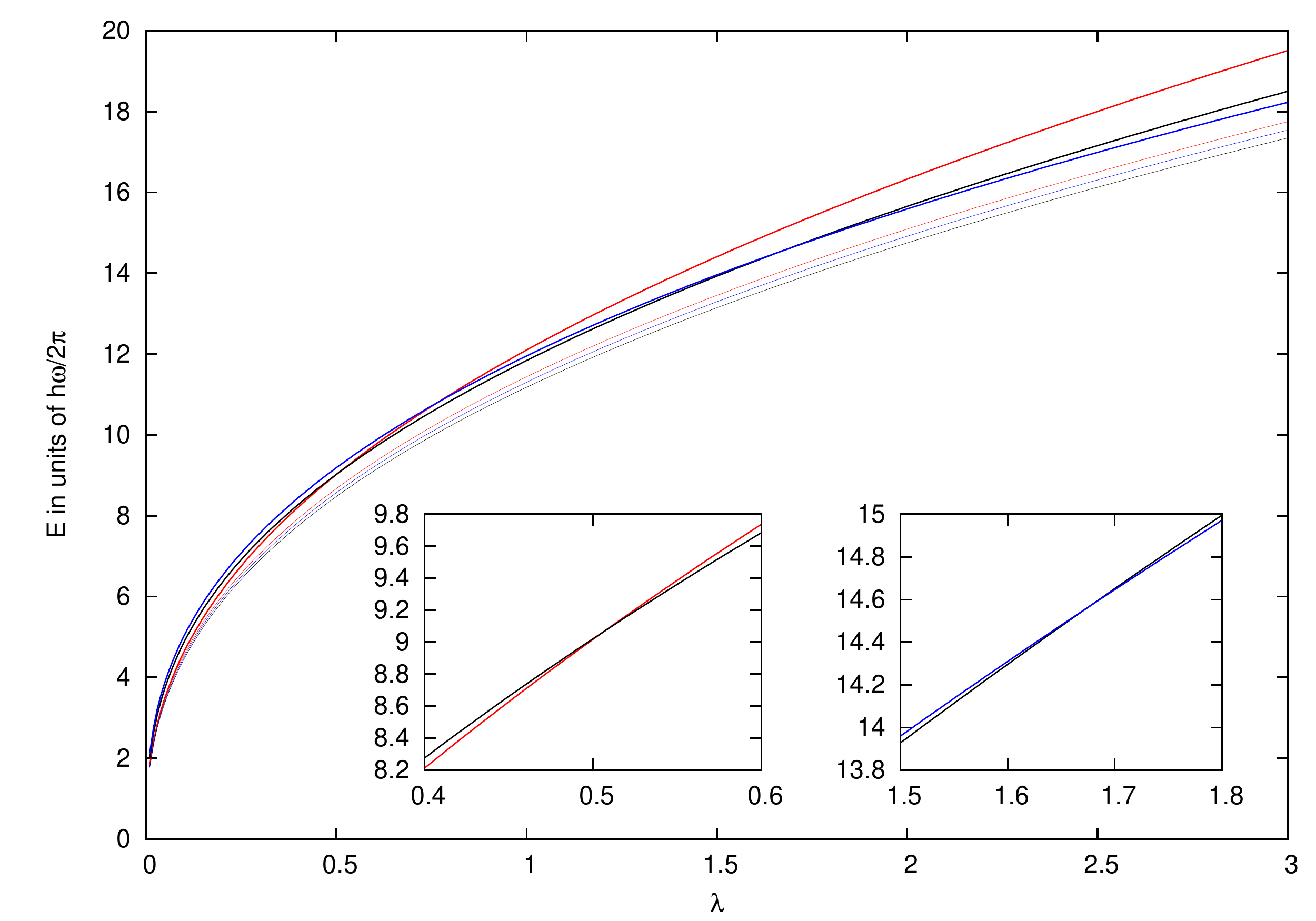}
\caption{\label{fig.E_lambda}
         The upper and lower set of three plots show the variation of $E$ 
         with and without the interface energy in TBEC of $^{85}$Rb-$^{87}$Rb
				 with parameter {\em set a}. Where $E$ of planar 
         (red curve), ellipsoidal (black curve) and cylindrical (blue curve) 
         geometries are examined as function of $\lambda$. The inset plots 
         show the region around the cross over points.
        }
\end{figure}

In the TFA calculations discussed so far, as mentioned earlier, the interface 
energy is neglected. Accordingly, the variational schemes we have adopted 
incorporate appropriate interface geometries. However, a general minimization 
by considering all the possible interface geometries favors the ellipsoidal 
interface as the equilibrium configuration. For example, though 
the cylindrical interface for $\lambda_i=50.0$ reproduces the numerical
results for TBEC of $^{85}$Rb-$^{87}$Rb with parameter {\em set a}, the 
minimization with ellipsoidal interface has lower $ E$. This is evident from 
the values of $E$, calculated over a wide range of $\lambda$ for 
the three interface geometries, shown in Fig. \ref{fig.E_lambda}. 

  As discussed in ref.\cite{Schaeybroeck}, the planer interface observed in
experiments \cite{Papp} emerges as the equilibrium geometry when the 
interface energy is considered. When $a_{\rm osc}$ is much larger than the 
interface thickness, the total excess energy arising from the finite interface 
tension \cite{Schaeybroeck} is
\begin{equation}
  \Omega_A = \frac{\sqrt{2 m_1}}{4\pi\hbar a_{11}}F(\xi_2/\xi_1,K)
             \int_A d\textbf{r}\left [ \mu_1-V(\textbf{r})\right ]^{3/2}.
   \label{int_energy}
\end{equation}
Here $\xi_i $ are the coherence lengths and 
$K$, $\xi_2/\xi_1$, $F(\xi_2/\xi_1,K)$ are defined as
\begin{eqnarray}
   K & = &\frac{(m_1+m_2)a_{12}}{2\sqrt{m_1m_2a_{11}a_{22}}},\nonumber\\
           \frac{\xi_2}{\xi_1} 
     & = & \left(\frac{m_1a_{11}}{m_2a_{22}}\right)^{1/4}, \nonumber\\
           F(\xi_2/\xi_1,K) & = & \frac{\sqrt{2}}{3}\left(1+
           \frac{\xi_2}{\xi_1}\right)- \frac{0.514\sqrt{\xi_2/\xi_1}}{K^{1/4}}-
                        \nonumber\\
     &   & \sqrt{\frac{\xi_2}{\xi_1}}\left(\frac{\xi_2}{\xi_1} 
           +\frac{\xi_1}{\xi_2}\right)\left(\frac{0.055}{K^{3/4}}+
           \frac{0.067}{K^{5/4}}\right)+\ldots . \nonumber
\end{eqnarray}
Here the integration is over the interface surface area. The above expression
is valid provided $K \ge 1.5$ and $ \xi_2/\xi_1\le1$. In the present work we 
consider TBECs in strongly segregated regime with $\xi_2/\xi_1<1$ and hence 
the interface energy in Eq.(\ref{int_energy}) is applicable. Then the 
interpenetration depth is proportional to $\sqrt{\xi_2\xi_1}/K^{1/4}$ and 
$\rightarrow 0 $ in the limit $1/K\rightarrow 0$. In this limit there 
is no overlap and TFA solution is an excellent approximation. The 
equilibrium geometry is then the one which minimizes the total energy: sum 
of TFA energy and $\Omega_A$. 

\begin{figure}[h]
\includegraphics[height=8cm,angle=-90]{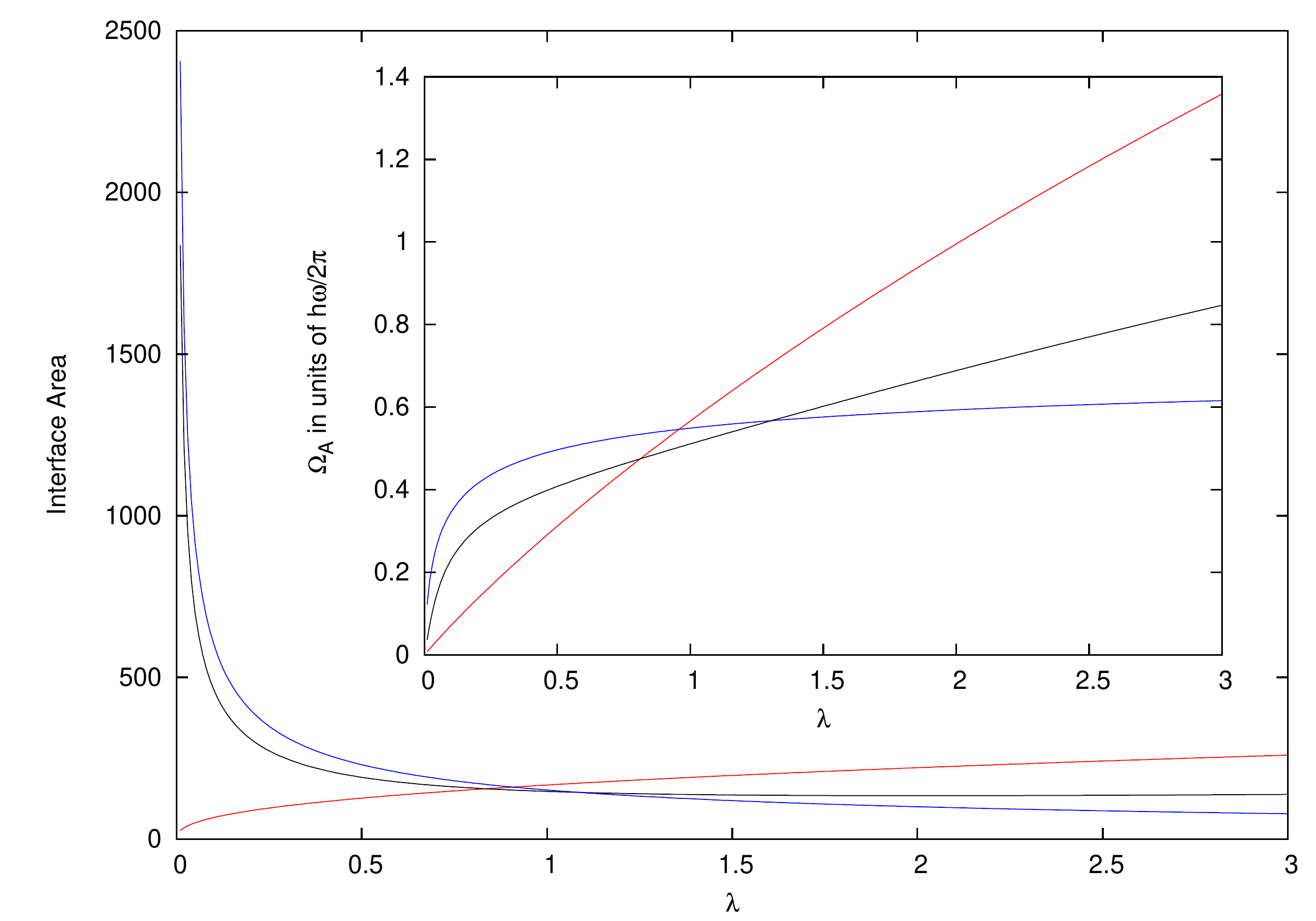}
\caption{\label{fig.Int_Areas} 
        Plots showing the interface areas as the functions of 
        $\lambda$, in the TBEC of $^{85}$Rb-$^{87}$Rb mixture, 
				for three geometries: planar (red curve), ellipsoidal 
        (black curve) and cylindrical (blue curve). The inset plot shows
        the interface energy.
        }
\end{figure}

 A precise determination of $\Omega_A$ is essential to obtain correct 
geometry of the phase separated TBEC. To a very good approximation, the 
interface energy is proportional to interface area. The interface area in 
planar and cylindrical geometries are 
$2\pi((2\mu_1-\lambda_1^2L_1^2)/\alpha_1^2)$  and
$4\pi\rho\sqrt{2\mu_1-\alpha_1^2\rho^2}/\lambda_1$ respectively. For 
prolate and oblate geometries the interface areas are
$2\pi R_1^2+2\pi R_1^2(\alpha_1\sin^{-1}e)/(e\lambda_1)$  and
and  $ 2\pi R_1^2+\pi(\alpha_1 R_1/\lambda_1)^2\ln((1+e)/(1-e))/e$ 
respectively. Here the ellipticity $e$ are $\sqrt{1-(\lambda_1/\alpha_1)^2}$ 
and $\sqrt{1-(\alpha_1/\lambda_1)^2}$ for prolate and oblate  respectively.
The interface areas in the three geometries for TBEC of $^{85}$Rb-$^{87}$Rb 
with parameter {\em set a}, using our semi analytic scheme developed in previous 
section, are shown in Fig. \ref{fig.Int_Areas}. The comparative study reveals 
that for $\lambda\ll1$, planar and ellipsoidal geometries have lower 
interface area than the cylindrical one. Whereas for $\lambda\gg1$, the 
cylindrical and ellipsoidal geometries have lower interface areas. In these 
two domains the interface area of one geometry is much lower than the other 
two and hence interface area can decide the preferred ground state geometry. 
For $\lambda$ close 
to one, the difference in the interface areas of the three geometries is 
small and surface tension is more crucial than interface area to 
determine the ground state geometry.

 In the following subsections we examine the impact of $\Omega_A $ in two 
domains: prolate shaped potentials ($\lambda_i<1$ ) and oblate shaped 
potentials ($\lambda_i>1$). For higher symmetry and simplified analysis 
we choose $\lambda_1=\lambda_2=\lambda$.


\subsection{Prolate trapping potentials}
  In the $\lambda<1$ domain, at low values of $\lambda$, the ellipsoidal 
geometry has higher ground state energy than the planar geometry. As 
$\lambda$ is increased, keeping the other parameters fixed, the ground state 
energies of both the geometries increase.  However, the planar geometry has
higher rate of increase. Then at $\lambda_a$, which is close to one, the  
energies of the two geometries are equal. Beyond this critical value, the 
energy of ellipsoidal geometry is lower and is the ground state geometry. 

\begin{figure}[h]
\includegraphics[height=8cm,angle=-90]{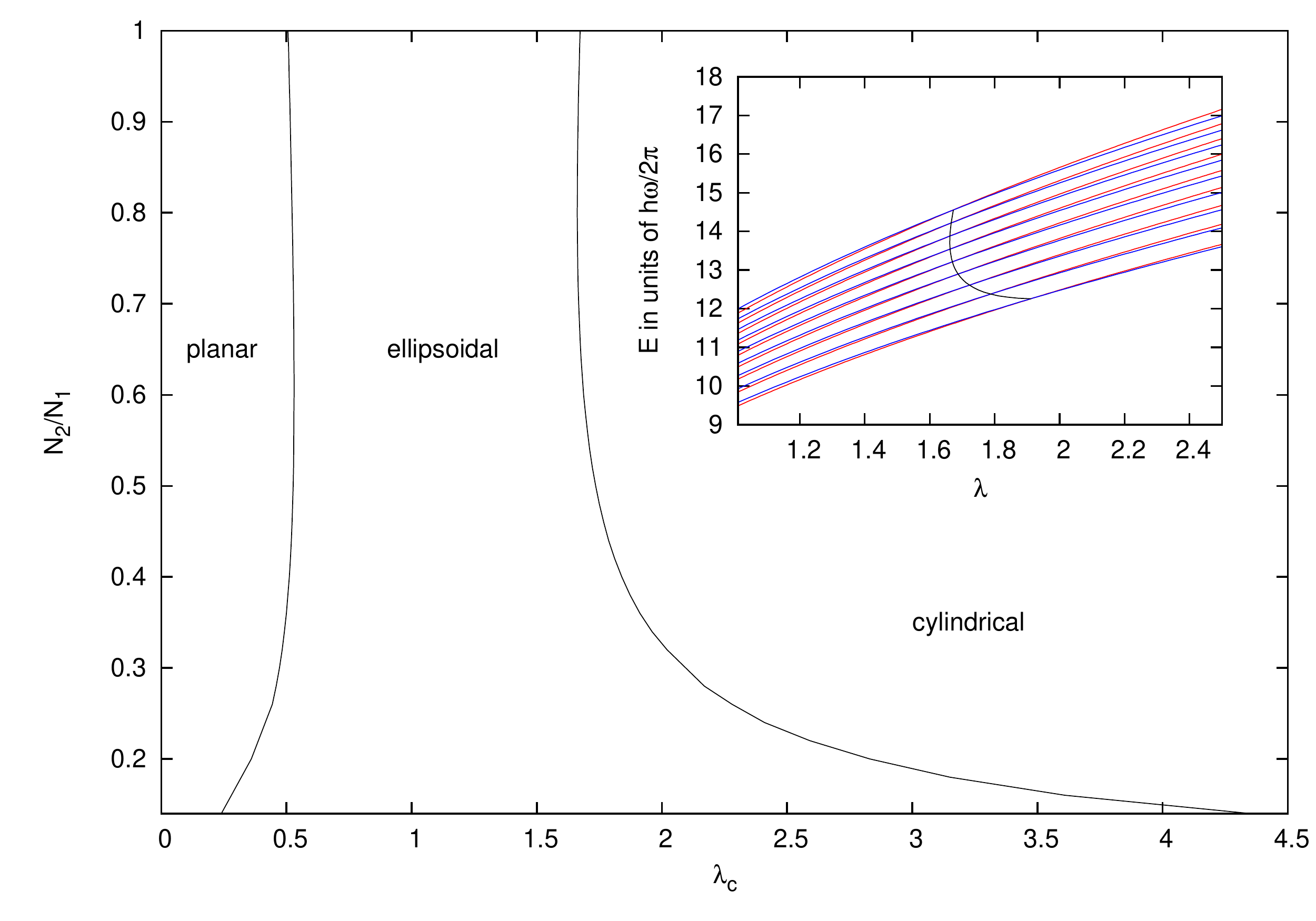}
\caption{\label{fig.phase_plot} Ground state geometry as a function of 
         $\lambda_c$ ($\lambda_a$ and $\lambda_b$ for planar-ellipsoidal and
         ellipsoidal-cylindrical transition respectively) and ratio of
         population $N_2/N_1$ in the TBEC of $^{85}$Rb-$^{87}$Rb mixture. 
         Inset plots show the variation in energy of two lowest energy 
         structures with $\lambda$ for oblate trapping potentials with 
         interface energy correction. Red and blue curves correspond to 
         ellipsoidal and cylindrical geometries respectively. Each pair 
         curves corresponds to different $N_2$ but same $N_1$. The uppermost 
         pair is for $N_1=N_2=50,000$. The next lower pair of curves has 
         $N_2=46,000,42,000$ and so on.
        }
\end{figure}

For  the $^{85}$Rb-$^{87}$Rb mixture with the parameter {\em set a},
the total energy $E$ and interface energy $\Omega_A$ as functions of 
$\lambda$ for the two geometries are respectively shown in 
Fig. \ref{fig.E_lambda} and Fig. \ref{fig.Int_Areas} (inset plot). 
Since the value of $\lambda_a$ depends on the parameters of the system, we 
examine the variation in $\lambda_a$ as function of the ratio $N_2/N_1$. 
For this we fix $N_1$ and vary $N_2$. Then calculate $\lambda_a$ as a 
function $N_2/N_1$. When $N_2$ is decreased $\lambda_a$ increases 
initially and then decreases. This is shown in 
Fig. \ref{fig.phase_plot}.


\subsection{Oblate trapping potentials}
 In the $\lambda>1$ regime, the ellipsoidal or cylindrical 
interface geometry is the preferred ground state geometry. The planar 
interface has higher $\Omega_A $ and not favored. For $\lambda$ 
close to $1$, the ellipsoidal geometry has lower energy, but looses 
stability as $\lambda$ is increased. This is due to the higher 
rate of increase in the $\Omega_A $ for ellipsoidal geometry. At the 
critical value  $\lambda_b$ and beyond, cylindrical geometry has lower 
total energy and takes over as the ground state geometry. 

For  the $^{85}$Rb-$^{87}$Rb mixture with the parameter {\em set a},
the total energy $E$ and interface energy $\Omega_A$ as functions of 
$\lambda$ for the two geometries are shown respectively in 
Fig. \ref{fig.E_lambda} and Fig. \ref{fig.Int_Areas} (inset plot).
For the same parameter {\em set a}; the value of $\lambda_b$, 
first decreases and then increases on decreasing $N_2$. This is shown in 
Fig. \ref{fig.phase_plot}.


\section{Conclusions}
 There are three distinct interface geometries of the ground state of TBEC in 
phase separated regime. We have developed a semi-analytic scheme to determine
the stationary state parameters for each of these and demonstrate the 
validity by comparing with the numerical results. We find in TFA, when the 
interface energy is neglected, the ellipsoidal geometry has the lowest 
energy for all values of $\lambda$. Hence is the preferred ground state 
structure. In this structure one species envelopes the other and interface 
geometry and overall shape of the TBEC is ellipsoidal. To explain the 
experimentally realized stationary state structures of TBECs we include the 
interface surface tension. We find that minimizing total energy, sum of 
TFA energy and $\Omega_A$, gives the right interface geometry. Then, in our 
semi-analytic scheme the ellipsoidal geometry no longer has the lowest energy 
for all values of $\lambda$. For cigar shaped traps ($\lambda \ll 1 $ ), the 
structure with the planar interface is the ground state geometry. While for pan 
cake shaped traps ($\lambda \gg 1 $ ) the cylindrical interface is the 
ground state geometry. For the values of $\lambda$ close to unity 
ellipsoidal structure is the ground state geometry.

\section{Appendix}
In case of planar interface between the two species trapped in potentials 
with separated minima, the expressions for $N_1$, $N_2$ and $E$ are:

\begin{eqnarray}
N_1 & = & \frac{\pi}{60U_{11}\alpha_1^2}\bigg(3\omega^2(l_1^5+L_1^5)m_1
          \lambda_1^4-20(l_1^3+L_1^3)\lambda_1^2\mu_1\nonumber\\
    &   & \frac{60 \left(l_1+L_1\right) \mu _1^2}{\omega ^2 m_1}\bigg),
\end{eqnarray}
\begin{eqnarray}
N_2 & = & \frac{1}{3\lambda _2 U_{22}}\pi\Bigg[\frac{8 \sqrt{2}\mu _2
          \left(\frac{\mu _2}{\omega ^2 m_2}\right)^{3/2}}{5 \alpha _2^2} + 
          \frac{1}{20 \omega ^2 m_2 \alpha _2^2}\nonumber\\
    &   & \left(5\omega ^4 l_1^5 m_2^2 \lambda _2^5-
		      \omega ^4 l_1^4 m_2^2 \lambda _2^4 \left(15 z_0 \lambda_2 + 
          8 l_1 \lambda _2\right)-\right.\nonumber\\
    &   & 10 z_0  \omega ^2 l_1^2 m_2 \lambda _2^3 
          \left(z_0 ^2 \omega ^2 m_2 \lambda _2^2-6 \mu _2\right)+ 
          20 z_0 \lambda_2 \nonumber\\
    &   & \left(z_0^2 \omega ^2 m_2 \lambda _2^2-3 \mu _2\right) \mu _2+
          60 \lambda_2 l_1 \left(z_0^2 \omega ^2 m_2 \lambda _2^2-
					\mu _2\right)\nonumber\\
    &   & \left. \mu _2+10 \omega ^2 l_1^3 m_2 \lambda _2^3 
          \left(-3 z_0^2 \omega ^2 m_2 \lambda _2^2 + 
					2 \mu _2\right)\right)\Bigg]\nonumber\\
    &   & -\frac{1}{3\lambda _2 U_{22}}\pi\Bigg[-\frac{8 \sqrt{2}\mu _2
           \left(\frac{\mu _2}{\omega ^2 m_2}\right)^{3/2}}{5 \alpha _2^2} + 
           \frac{1}{20 \omega ^2 m_2 \alpha _2^2}\nonumber\\
    &   & \left(-5\omega ^4 L_1^5 m_2^2 \lambda _2^5+
          \omega ^4 L_1^4 m_2^2 \lambda _2^4 \left(-15 z_0 \lambda_2 + 
          8 L_1 \lambda _2\right)-\right.\nonumber\\
    &   & 10 z_0  \omega ^2 L_1^2 m_2 \lambda _2^3 
          \left(z_0^2 \omega ^2 m_2 \lambda _2^2-6 \mu _2\right)+ 
         20 z_0 \lambda_2 \nonumber\\
    &   & \left(z_0 ^2 \omega ^2 m_2 \lambda _2^2-3 \mu _2\right) \mu _2-
          60 \lambda_2 L_1 \left(z_0 ^2 \omega ^2 m_2 \lambda _2^2-
				  \mu _2\right)\nonumber\\
    &   & \left. \mu _2+10 \omega ^2 L_1^3 m_2 \lambda _2^3 
          \left(3 z_0 ^2 \omega ^2 m_2 \lambda _2^2 - 2 \mu _2\right)\right)
					\Bigg],
\end{eqnarray}
\begin{eqnarray}
E & = & \frac{1}{60 U_{22}}\pi \Bigg[\frac{160 \sqrt{2} \mu_2^3 
        \sqrt{\frac{\mu_2}{\omega^2 \lambda_2^2 m_2}}}{7 \omega^2 m_2 
				\alpha_2^2}-\frac{1}{14 \omega^2 m_2 \alpha_2^2}\nonumber\\
  &   & \left(21 \omega^6 l_1^7 m_2^3 \lambda_2^6+\omega^6 l_1^6 m_2^3 
	      \lambda_2^6 \left(35 z_0 -16	l_1\right)+\right.\nonumber\\
	&   & 7 \omega^4 l_1^5 m_2^2 \lambda_2^4 \left(15 z_0^2 \omega^2 m_2 
	      \lambda_2^2-	16 \mu_2\right)+7\omega^4 l_1^4 m_2^2 \lambda_2^4 
				\nonumber\\
	&   & \left(25 z_0 ^3 \omega ^2 m_2 \lambda _2^2+16 l_1 \mu _2\right)+
	      7 z_0  \omega ^2 l_1^2 m_2 \lambda _2^2\nonumber\\
	&   & \left(3 z_0 ^4 \omega ^4 m_2^2 \lambda _2^4-40 z_0 ^2 \omega ^2 m_2 
	      \lambda _2^2 \mu _2-60 \mu _2^2\right)-\nonumber\\
	&   & 14 z_0  \mu _2 \left(3 z_0 ^4 \omega ^4 m_2^2 \lambda _2^4+
	      10 z_0 ^2 \omega ^2 m_2 \lambda _2^2 \mu _2-40 \mu _2^2\right)
				\nonumber\\
	&   & +70 l_1 \mu _2 \left(-3 z_0 ^4 \omega ^4 m_2^2 \lambda _2^4-
	      6 z_0 ^2 \omega ^2 m_2 \lambda _2^2 \mu _2+8 \mu _2^2\right)
				\nonumber\\
	&   & \left.+35 l_1^3 \left(3 z_0 ^4 \omega ^6 m_2^3 \lambda _2^6-4 
	      \omega ^2 m_2 \lambda _2^2 \mu _2^2\right)\right)\Bigg]\nonumber\\
  &   & +\frac{1}{60 U_{22}}\pi \Bigg[\frac{160 \sqrt{2} \mu _2^3 
	      \sqrt{\frac{\mu _2}{\omega ^2 \lambda _2^2 m_2}}}{7 \omega ^2 m_2 
				\alpha _2^2}+\frac{1}{14 \omega ^2 m_2 \alpha _2^2}\nonumber\\
  &   & \left(-21 \omega ^6 L_1^7 m_2^3 \lambda _2^6+\omega ^6 L_1^6 m_2^3 
	      \lambda _2^6 \left(35 z_0 +16	L_1\right)-\right.\nonumber\\
	&   & 7 \omega ^4 L_1^5 m_2^2 \lambda _2^4 \left(15 z_0 ^2 \omega ^2 m_2 
	      \lambda _2^2-	16 \mu _2\right)+7 \omega ^4 L_1^4 m_2^2 \lambda _2^4
				\nonumber\\
	&   & \left(25 z_0 ^3 \omega ^2 m_2 \lambda _2^2-16 L_1 \mu _2\right)+
	      7 z_0  \omega ^2 L_1^2 m_2 \lambda _2^2\nonumber\\
	&   & \left(3 z_0 ^4 \omega ^4 m_2^2 \lambda _2^4-40 z_0 ^2 \omega ^2 m_2 
	      \lambda _2^2 \mu _2-60 \mu _2^2\right)-\nonumber\\
	&   & 14 z_0  \mu _2 \left(3 z_0 ^4 \omega ^4 m_2^2 \lambda _2^4+
	      10 z_0 ^2 \omega ^2 m_2 \lambda _2^2 \mu _2-40 \mu _2^2\right)
				\nonumber\\
	&   & +70 L_1 \mu _2 \left(3 z_0 ^4 \omega ^4 m_2^2 \lambda _2^4+6 z_0 ^2 
	      \omega ^2 m_2 \lambda _2^2 \mu _2-8 \mu _2^2\right)\nonumber\\
	&   & \left.-35 L_1^3 \left(3 z_0 ^4 \omega ^6 m_2^3 \lambda _2^6-4 
	       \omega ^2 m_2 \lambda _2^2 \mu _2^2\right)\right)\Bigg]\nonumber\\
  &   & +\frac{\pi}{168 U_{11}\alpha_1^2} \bigg(\omega ^4 \left(L_1^7+L_1^7
	      \right) m_1^2 \lambda _1^6-28 \left(L_1^3+L_1^3\right) \lambda _1^2 
				\mu _1^2+ \nonumber\\
  &   & \frac{112 \left(L_1+L_1\right) \mu _1^3}{\omega ^2 m_1}\bigg).
\end{eqnarray}
These equations reduce to those for coincident centers on substituting 
$l_1=L_1$ and $z_0=0$.



\begin{thebibliography}{99}
  \bibitem{Myatt}
     C.~J.~Myatt, E.~A.~Burt, R.~W.~Ghrist, E.~A.~Cornell, and C.~E.~Wieman,
     Phys. Rev. Lett. {\bf 78}, 586 (1997).
  \bibitem{Modugno}
     G.~Modugno, M.~Modugno, F.~Riboli, G.~Roati, and M.~Inguscio,
     Phys. Rev. Lett. {\bf 89}, 190404 (2002).
  \bibitem{Papp}
     S.~B.~Papp, J.~M.~Pino and C.~E.~Wieman,
     Phys. Rev. Lett. {\bf 101}, 040402 (2008).
  \bibitem{Gautam}
     S.~Gautam, and D.~Angom,
     arXiv:0908.4336v3.
  \bibitem{Sasaki}
     K.~Sasaki, N.~Suzuki, D.~Akamatsu, and H.~Saito,
     arXiv:0910.1440v1.
  \bibitem{Takeuchi}
     H.~Takeuchi, N.~Suzuki, K.~Kasamatsu, H.~Saito, and M.~Tsubota,
     arXiv:0909.2144v1. 
  \bibitem{Saito}
     H.~Saito, Y.~Kawaguchi, M.~Ueda,
     Phys. Rev. Lett. {\bf 102}, 230403 (2009).
  \bibitem{Ho}
     Tin-Lun Ho, and V.~ B.~Shenoy,
     Phys. Rev. Lett. {\bf 77}, 3276 (1996).
  \bibitem{Timmermans}
     E.~Timmermans,
     Phys. Rev. Lett. {\bf 81}, 5718 (1998).
  \bibitem{Ao}
     P.~Ao, and S.~T.~Chui,
     Phys. Rev. A {\bf 58}, 4836 (1998).
  \bibitem{Barankov}
     R.~A.~Barankov,
     Phys. Rev. A {\bf 66}, 013612 (2002).
  \bibitem{Schaeybroeck}
     B.~Van Schaeybroeck,
     Phys. Rev. A {\bf 78}, 023624 (2008).
  \bibitem{Trippenbach}
     M.~Trippenbach, K.~Goral, K.~Rzazewski, B.~Malomed, and Y.~B.~Band
     J. Phys. B {\bf 33}, 4017 (2000).
  \bibitem{Muruganandam}
     P.~Muruganandam, and S.~K.~Adhikari,
     Comp. Phys. Comm. {\bf 180}, 1888 (2009).
 \end{thebibliography}
\end{document}